\documentclass[prd,aps,preprint,draft,showpacs]{revtex4}
\usepackage{epsf}
\usepackage[dvips]{color}
\usepackage{dcolumn}
\newcommand{\Tr}{\mbox{Tr}}
\newcommand{\VEV}[1]{\left\langle #1\right\rangle}

\begin{document}
\title{Topological susceptibility in staggered fermion chiral perturbation theory}
\author{Brian Billeter, Carleton DeTar, and James Osborn}
\affiliation{
Physics Department, University of Utah, Salt Lake City, UT 84112, USA
}
\date{\today}

\begin{abstract}
The topological susceptibility of the vacuum in quantum chromodynamics
has been simulated numerically using the Asqtad improved staggered
fermion formalism.  At nonzero lattice spacing the residual fermion
doublers (fermion ``tastes'') in the staggered fermion formalism give
contributions to the susceptibility that deviate from conventional
continuum chiral perturbation theory.  In this brief report we
estimate the taste-breaking artifact and compare it with results of
recent simulations, finding that it accounts for roughly half of the
scaling violation.
\end{abstract}

\maketitle

\section{Introduction}

The staggered fermion formalism reduces the number of fermion doublers
to four species (``tastes'') per quark flavor.  At zero quark mass in
the continuum limit the taste symmetry group becomes
SU(4)$_L\times$SU(4)$_R\times$U(1)$_V$, but at nonzero lattice spacing
this symmetry is broken by lattice-artifact, taste-changing
interactions.  Taste breaking effects modify predictions of chiral
perturbation theory, complicating efforts to extrapolate lattice
results simultaneously to zero quark mass and zero lattice spacing.
However, a recently developed modification to chiral perturbation
theory, called ``staggered chiral perturbation theory (S$\chi$PT)'',
permits a quantitative treatment of taste-breaking effects
\cite{ref:lee.sharpe,ref:aubin.bernard}.  Combining these
modifications with an improved staggered fermion formulation that
substantially reduces the strength of taste breaking has proven highly
effective in carrying out high-precision extrapolations of results
from lattice simulations \cite{ref:precision}.

In this brief report we use staggered chiral perturbation theory to
predict the topological susceptibility.  The derivation is given in
Sec.~\ref{sec:formalism} and in Sec.~\ref{sec:compare} the resulting
formula is compared with results of recent lattice simulations with
improved staggered fermions.

\section{Topological Susceptibility with Taste Symmetry Breaking}
\label{sec:formalism}

We consider the low energy effective chiral Lagrangian for $N$ flavors
of staggered fermions, each with four tastes.  The Lagrangian is
written in terms of the $(4N)^2$ pseudoscalar meson fields $\phi_{ia}$,
encapsulated in a unitary $4N \times 4N$ matrix
\begin{equation}
  U = \exp(i T_{ia} \phi_{ia}/f).
\end{equation}
We use $f \approx 130$ MeV.  We label the generators of the Lie
algebra of U(4N) with a flavor index $i$ and taste index $a$ and
write them as
\begin{equation}
  T_{ia} = \Lambda_i t_a,
\end{equation}
where the $t_a$, the 16 taste matrices (Dirac gamma matrices plus
the identity), are enumerated as
\begin{equation}
  \xi_5, \xi_\mu, i\xi_{\mu5} = i\xi_\mu \xi_5 , 
     i \xi_{\mu\nu} = i \xi_\mu \xi_\nu, \xi_I = I
\end{equation}
and the $\Lambda_i$ are the $N^2$ generators of U(N).  The $t_a$ are
orthonormal:
\begin{equation}
   \Tr(t_a t_b ) = 4 \delta_{ab},
\end{equation}
but we normalize the $\Lambda_i$ so that
\begin{equation}
   \Tr(\Lambda_{i} \Lambda_{j} ) = \delta_{ij}.
\end{equation}
Moreover, for convenience we choose a quark flavor basis in which
the diagonal generators are enumerated first and are explicitly
\begin{equation}
   (\Lambda_i)_{jk} =  \delta_{ij} \delta_{ik}
    \ \ \ \mbox{for $i = 1,\ldots,N$}.
\end{equation}
Also, for convenience, we define the flavor-singlet generator and fields
\begin{eqnarray}
  \Lambda_0 &=& I \\
  \phi_{0a} &=& \sum_{i = 1}^N \phi_{ia}/\sqrt{N}.
\end{eqnarray}

The Euclidean Lee-Sharpe S$\chi$PT Lagrangian \cite{ref:lee.sharpe},
extended by Aubin and Bernard to $N$ flavors \cite{ref:aubin.bernard},
and including an explicit mass term for the anomalous flavor singlet
field is
\begin{equation}
  {\cal L} = 
   \frac{f^2}{8} \Tr \left(\partial_\mu U^\dagger \partial_\mu U \right)
    - \frac{\mu f^2}{4} \Tr[{\cal M}(U^\dagger + U)] 
   + {m_0^2 \over 2}\phi_{0I}^2
   + a^2 {\cal V}(U)
\end{equation}
where $a$ is the lattice spacing, the trace is over all $4N$ taste and
flavor indices, ${\cal M}$ is the taste-degenerate quark mass matrix
\begin{equation}
  {\cal M}_{ia,jb} = M_{ij}\delta_{ab} = \delta_{ij} \delta_{ab} m_i,
\end{equation}
and the taste-breaking potential,
\begin{equation}
  -{\cal V}(U) = \sum C_i {\cal O}_i,
\end{equation}
is a linear combination of the operators
\begin{eqnarray}
  {\cal O}_1 &=& \Tr\left( T_{0,5} U T_{0,5} U^\dagger \right) \\
  {\cal O}_{2V} &=& {1 \over 4} \left[
      \Tr( T_{0,\mu} U ) \Tr( T_{0,\mu} U ) + \mbox{h.c.}\right] \\
  {\cal O}_{2A} &=& {1 \over 4} \left[
      \Tr( T_{0,\mu 5} U ) \Tr( T_{0,5\mu} U ) + \mbox{h.c.}\right] \\
  {\cal O}_3 &=& {1 \over 2}\left[
      \Tr( T_{0,\mu} U T_{0,\mu} U ) + \mbox{h.c.}\right] \\
  {\cal O}_4 &=& {1 \over 2}\left[
      \Tr( T_{0,\mu 5} U T_{0,5\mu} U ) + \mbox{h.c.}\right] \\
  {\cal O}_{5V} &=& {1 \over 2} \left[
      \Tr( T_{0,\mu} U ) \Tr( T_{0,\mu} U^\dagger )\right] \\
  {\cal O}_{5A} &=& {1 \over 2} \left[
      \Tr( T_{0,\mu 5} U ) \Tr( T_{0,5 \mu} U^\dagger )\right] \\
  {\cal O}_6 &=& \sum_{\mu < \nu}
       \Tr\left( T_{0,\mu\nu} U T_{0,\nu\mu} U^\dagger \right)  \ \ .
\end{eqnarray}
The partition function is, as usual
\begin{equation}
   Z = \int [dU] \exp(-{\int \cal L} dV).
\end{equation}
The chiral condensate is defined by
\begin{equation}
  \Sigma_i = 
    \lim_{m_i \rightarrow 0}{1 \over 4} \frac{\partial \log Z}{\partial m_i}
\end{equation}
and is, as usual,
\begin{equation}
  \Sigma_i = \Sigma = \mu f^2/2
\end{equation}

The bare meson masses are obtained by expanding the Lagrangian to quadratic
order in the meson fields:
\begin{equation}
  {\cal L} = \frac{1}{2} \sum_{ia} (\partial_\mu \phi_{ia})^2
  + \frac{1}{2} \sum_{iajb} A_{ia,jb} \phi_{ia} \phi_{jb}
\end{equation}
where the flavor and taste mixing matrix is
\begin{equation}
  A_{ia,jb} = A^{(0)}_{ij} \delta_{ab} + 
              A^{(1)}_{ij}\delta_{Ia}\delta_{Ib} +
              a^2 B_{ab}\delta_{ij}.
\end{equation}
The continuum taste-singlet flavor-mixing matrix is block diagonal
with an invariant subspace $i = 1,2,\ldots{},N$.  This is the only
subspace that contributes to the tree-level topological susceptibility.
For $i,j$ in this set we have
\begin{eqnarray}
  A^{(0)}_{ij} &=&  2 \mu \Tr(M \Lambda_i \Lambda_j) = 2\mu \delta_{ij}m_i \\
  A^{(1)}_{ij} &=&  m_0^2/N.
\end{eqnarray}
The flavor-singlet taste-mixing matrix $B_{ab}$ is defined by the
small field expansion
\begin{equation}
  {\cal V}(U) \approx {1 \over 2}\sum_{i,a,b} B_{ab} \phi_{ia}\phi_{ib}.
\end{equation}

Without the anomalous mass term $m_0^2$ and without taste breaking the
bare flavor-neutral meson masses are given by the diagonal values in $A^{(0)}$.
\begin{equation}
  m^2_{ia} = 2 \mu m_i \ \ \ \mbox{for $i = 1,2,\ldots{},N$.}
\end{equation}
%
To lowest order in the taste-breaking potential the pseudoscalar mass
splittings were calculated by Lee and Sharpe\cite{ref:lee.sharpe}.
With the Aubin and Bernard generalization\cite{ref:aubin.bernard} they
are
\begin{equation}
  \Delta m_{ia}^2 = a^2 \Delta_a,
\end{equation}
where $\Delta_a = B_{aa}$ and 
\begin{eqnarray}
  \Delta_5 &=& 0 \\
  \Delta_\mu &=& \frac{16}{f^2}(C_1 + C_3 + 3 C_4 + 3 C_6) \\
  \Delta_{\mu 5} &=& \frac{16}{f^2}(C_1 + 3 C_3 +  C_4 + 3 C_6) \\
  \Delta_{\mu \nu} &=& \frac{16}{f^2}(2 C_3 + 2 C_4 + 4 C_6) \\
  \Delta_I &=& \frac{16}{f^2}(4 C_3 + 4 C_4) \ \ .
\end{eqnarray}

To formulate the topological susceptibility in the presence of taste
breaking, we recall the derivation, starting from the quark and gluon
Lagrangian ${\cal L}_{QCD}$.  The partition function receives
contributions from topological charge sectors $\nu$ as follows
\cite{ref:LS1992}:
\begin{equation}
  Z = \sum_\nu Z_\nu \ \ \ ,
\end{equation}
where the microcanonical partition function is
\begin{equation}
  Z_\nu = \int dq \, d\bar q \, dA \, d\theta \exp(-i \theta \nu)
    \exp\left( \int [{\cal L}_{QCD} + i\theta F \tilde F/64\pi^2] dV \right).
\end{equation}
The theta term in the Lagrangian can be removed by a flavor (and
taste) singlet $U_A(1)$ chiral rotation, leaving a Lagrangian ${\cal
L}_{QCD}(\theta)$ with rotated quark fields.  The lattice staggered
fermion action breaks this symmetry through the quark mass term and
through taste symmetry breaking at nonzero lattice spacing $a$.  To
order $a^2$ in S$\chi$PT the symmetry breaking is described by the
$U_A(1)$ non-invariant taste-breaking term $a^2{\cal V}(U)$.  The
$U_A(1)$ rotation on the quark fields is modeled in the effective
theory by the corresponding rotation on the meson matrix $U$:
\begin{equation}
  {\cal L}(\theta) = 
   \frac{f^2}{8} \Tr \left(\partial_\mu U^\dagger \partial_\mu U \right)
    - \frac{\mu f^2}{4} \Tr[{\cal M}(e^{-i\theta/4N}U^\dagger + 
     e^{i\theta/4N}U)] 
   + {m_0^2 \over 2}\phi_{0I}^2
   + a^2 {\cal V}(e^{i\theta/4N}U).
\end{equation}
A field redefinition $\phi_{iI} \rightarrow \phi_{iI} - f\theta/4N$
for $i = 1,\ldots{}N$, implying $\phi_{0I} \rightarrow \phi_{0I} -
f\theta/4\sqrt{N}$, gives an alternate expression
\begin{equation}
  {\cal L}(\theta) = 
   \frac{f^2}{8} \Tr \left(\partial_\mu U^\dagger \partial_\mu U \right)
    - \frac{\mu f^2}{4} \Tr[{\cal M}(U^\dagger + U)] 
   + {m_0^2 \over 2}(\phi_{0I} - f\theta/4\sqrt{N})^2
   + a^2 {\cal V}(U).
\label{eq:chiraltheta}
\end{equation}

We work in the mean-field $\epsilon$ regime in which the fields are
space and time independent and take the limit $m \Sigma V \rightarrow
\infty$, and $m_0^2 V \rightarrow \infty$ i.e. asymptotically large
mean square topological charge $\VEV{\nu^2}$.  In this limit we may
use a saddle-point approximation to compute the microcanonical
partition function.  The saddle point occurs close to the identity in
$U$ and origin in $\theta$, where we replace the Haar measure $dU$
with the $(4N)^2$ dimensional Cartesian measure $d\phi$ and expand the
integrand to quadratic order in the meson field $\phi$ and $\theta$:
\begin{equation}
  Z_\nu \propto \int d\phi \, d\theta \exp(-i \theta \nu) 
       \exp\left\{-{V \over 2} \left[F(\phi) 
      + m_0^2 (\phi_{0I} - f\theta/4\sqrt{N})^2 \right] \right\}
\end{equation}
where $F(\phi)$ is the quadratic form
\begin{equation}
  F(\phi_{ia}) = \sum_{ia,jb} \left[ 
   A^{(0)}_{ij}\delta_{ab} +
   a^2 B_{ab}\delta_{ij}\right]\phi_{ia}\phi_{jb} .
\end{equation}
The only meson fields contributing to the susceptibility are the
diagonal taste-singlet mesons $\phi_{iI}$ for $i = 1,2,\ldots{},N$.
The corresponding subspace is invariant under the quadratic form.
Thus the remaining fields can be integrated out immediately. The
integration over theta gives
\begin{equation}
  Z_\nu \propto \int \prod_{i=1}^N d\phi_{iI} 
       \exp(-i 4\sqrt{N}\phi_{0I} \nu/f) 
       \exp\left[-{V \over 2} F(\phi_{iI}) - 16N \nu^2/(2 V f^2 m_0^2)\right]
\end{equation}
and finally the integration over taste-singlet meson fields gives
\begin{equation}
  Z_\nu \propto 
       \exp\left[- 16 \nu^2/\left(2 V f^2 \sum_{i=1}^N 1/m^2_{iI}\right) 
    - 16 N \nu^2/(2 V f^2 m_0^2) \right]
\end{equation}
where the bare taste-shifted flavor-neutral masses are
\begin{equation}
    m^2_{iI} = 2 \mu m_i + a^2 \Delta_I.
\end{equation}
The resulting topological susceptibility is
\begin{equation}
  \chi = \VEV{\nu^2}/V = 
     \frac{f^2/16}{\sum_{i=1}^N 1/m^2_{iI} + N/m_0^2}
\end{equation}
When the $N$ flavors can be grouped into $p$ degenerate flavor
multiplets with degeneracy $n_k$ so that $\sum_{k=1}^p n_k = N$, the
susceptibility is
\begin{equation}
  \chi = \VEV{\nu^2}/V = 
     \frac{f^2/16}{\sum_{k=1}^p n_k (1/m^2_{kI} + 1/m_0^2)}
\end{equation}

Counting the taste symmetry as well, the continuum limit degeneracy of
quark species is then $4n_k$ for each distinct flavor.  In numerical
simulations of staggered fermions it is necessary to take fractional
powers of the quark determinant to simulate a theory with the physical
count of $2+1$ flavors, {\it i.e.}, only two species of light $u$ and
$d$ quarks of the same mass and a single $s$ quark.  The S$\chi$PT
treats only the case in which quark multiplets come in multiples of
four.  Absent a rigorous treatment of fractional powers in S$\chi$PT,
we adopt the replica trick: assume that an analytic continuation in
the flavor number $n_k$ reproduces the effect of taking a fractional
power of the determinant.  That is, we replace $4n_{ud}$ by 2 and
$4n_s$ by 1 in the expression above.  The susceptibility for $2+1$
flavors is then
\begin{equation}
  \chi = 
     \frac{f^2 m^2_{\pi,I}/8}
      {1 +  m^2_{\pi,I}/2 m^2_{ss,I} + 3 m^2_{\pi,I}/2 m_0^2}
\label{eq:suscept_tb}
\end{equation}
Of course, this result is valid only in leading order chiral perturbation
theory, but at large quark mass it interpolates smoothly
\cite{ref:Durr} to the quenched result predicted for pure gauge theory
\cite{ref:witten.veneziano}.
\begin{equation}
   \chi_q = f^2 m_0^2/12.
\label{eq:quenched}
\end{equation}
At small $m_{ud}$ the susceptibility is dominated by the pion mass.
Without taste breaking we recover the conventional formula
\begin{equation}
  \chi = f^2 m^2_\pi/8.
\end{equation}
We see that the effect of taste breaking at leading order is simply to
replace the masses of the Goldstone mesons in the conventional
expression with the masses of the corresponding taste-singlet
non-Goldstone states.  The squared masses of these states vary
linearly with quark mass and have a nonzero intercept.  Therefore, at
nonzero lattice spacing, the topological susceptibility does not
vanish at zero quark mass.  However, in the continuum limit, the
expected zero is recovered.


\section{Comparison with Simulation Results}
\label{sec:compare}

The topological susceptibility has been determined recently for gauge
configurations generated with an improved staggered fermion
action\cite{ref:topo.susc}.  It was found that at nonzero lattice
spacing the lattice susceptibility is higher than that predicted by
chiral perturbation theory, but (at least within the available
statistical errors and an extrapolation based on limited data) it is
consistent with the continuum prediction.  Two principal causes for
disagreement at nonzero lattice spacing were proposed: (1)
Dislocations in the gluon field could imitate small instantons and
participate in screening of the topological charge.  These and small
instantons are erased in the smoothing (cooling) process, possibly
leaving behind larger, unscreened instantons.  The latter would then
increase the smoothed susceptibility.  (2) The would-be zero modes are
not at zero because of taste-symmetry breaking, so are not fully
effective in suppressing the topological charge at small quark mass.
The latter effect is taken into account in Eq (\ref{eq:suscept_tb}).

\begin{figure}[ht]
\begin{tabular}{ll}
 \epsfxsize=70mm
 \epsfbox{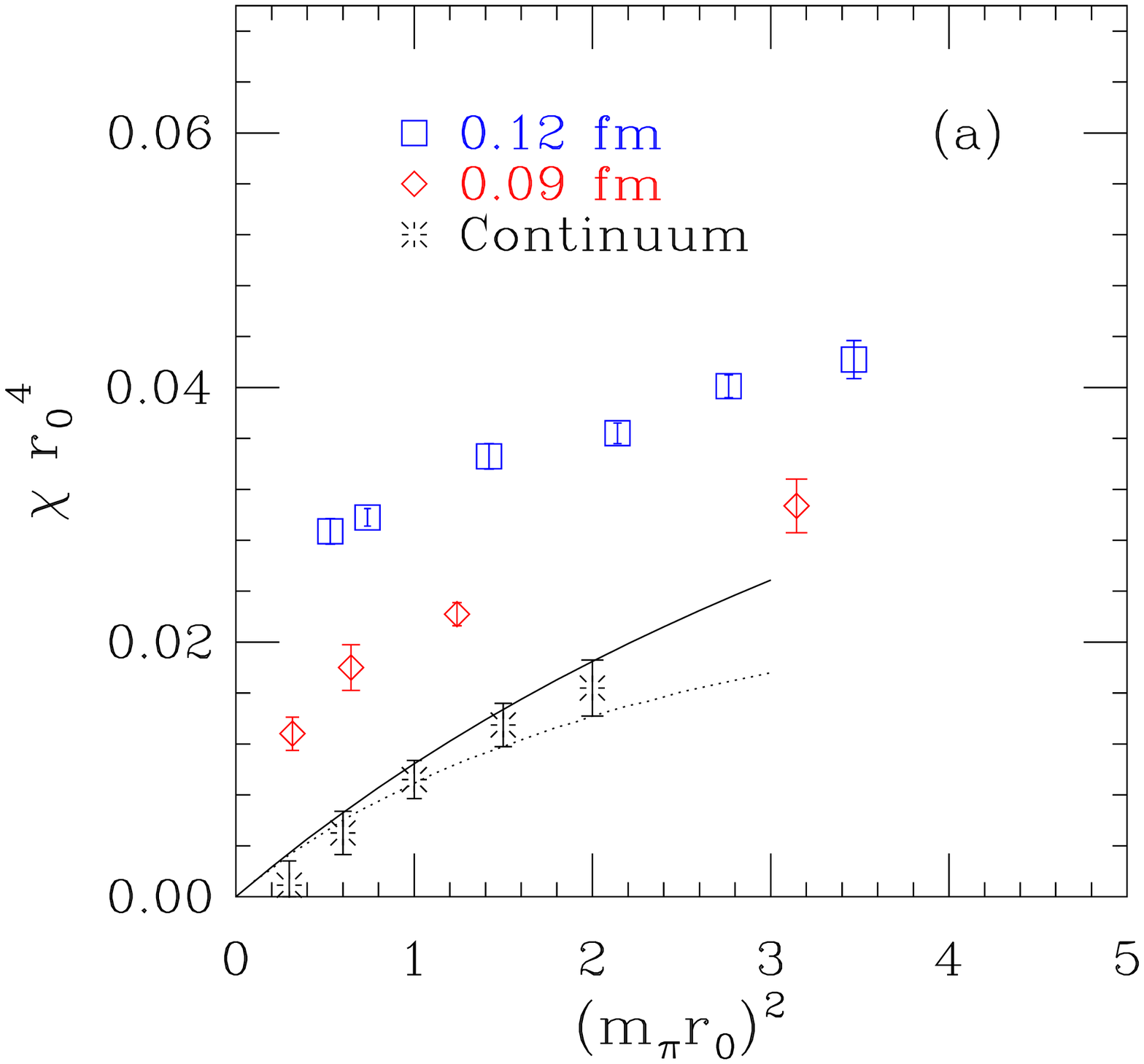}
&
 \epsfxsize=70mm
 \epsfbox{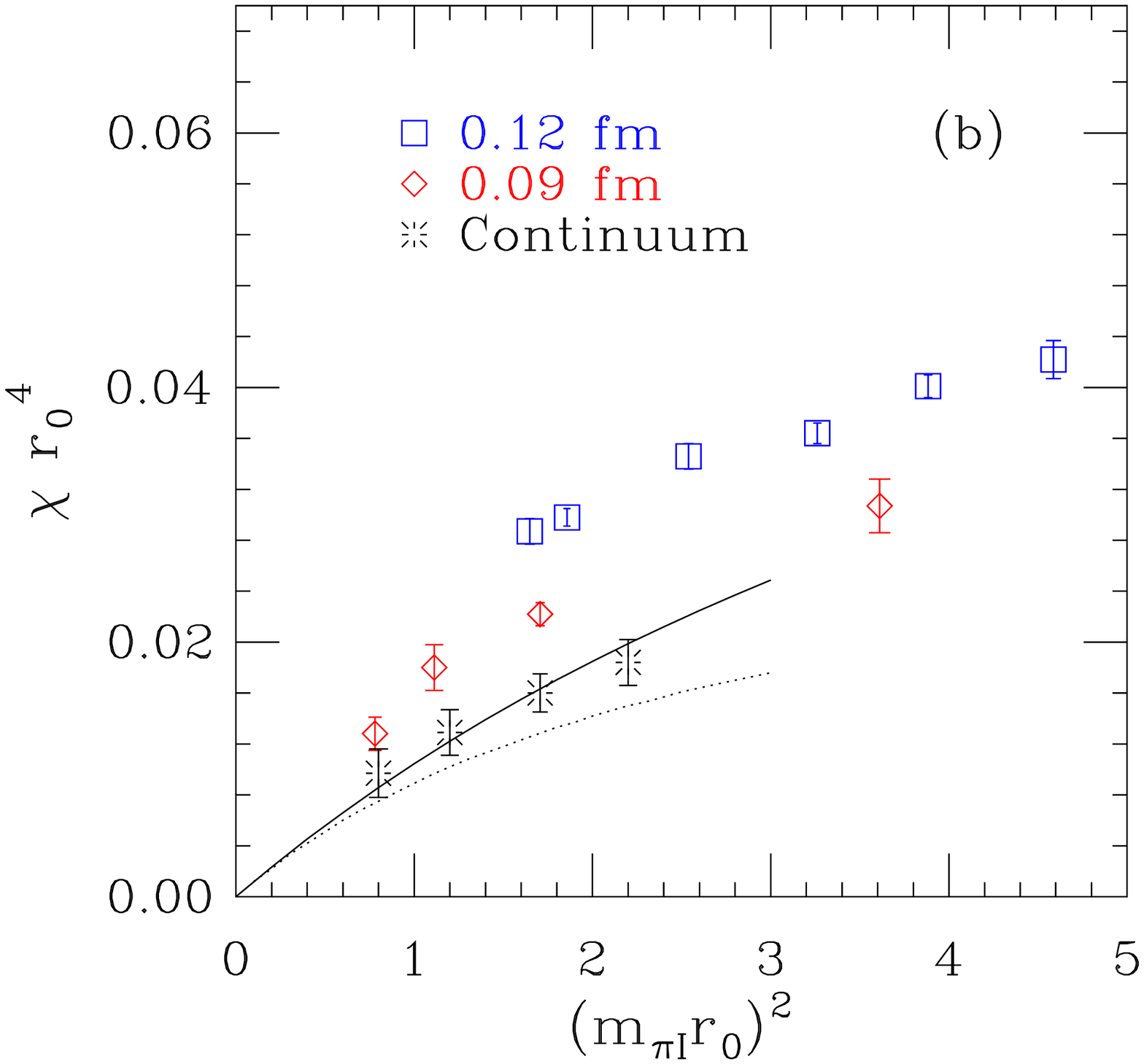}
\end{tabular}

\caption{Topological susceptibility {\it vs} pion mass squared in
units of $r_0$ with improved (Asqtad) staggered quarks.  Data is from
Ref.~\protect\cite{ref:topo.susc}, reanalyzed with a new variance
reduction method and new 0.09 fm data point
\protect\cite{ref:reducevariance}: (a) plotted as a function of the
Goldstone pion mass squared (b) replotted as a function of the
taste-singlet pion mass squared.  The bursts show the continuum
extrapolation at fixed abscissa in each case. The solid line shows the
prediction of leading order continuum chiral perturbation theory [Eq
(\protect\ref{eq:suscept_tb})] without the $U(1)$ mass term $m_0^2$.
The dashed line includes $m_0^2$, determined through Eq
(\ref{eq:quenched}) from the simulated quenched susceptibility
\protect\cite{ref:reducevariance}.
\label{fig:chi_vs_mpi2} }

\end{figure}

In light of our result [Eq (\ref{eq:suscept_tb}] we replot reanalyzed
data of Ref.~\cite{ref:topo.susc} as a function of the taste-singlet
pion mass $m_{\pi,I}^2 r_0^2$ in Fig.~\ref{fig:chi_vs_mpi2}(b). For
comparison the same data is plotted as a function of the Goldstone
pion mass $m_{\pi}^2 r_0^2$ in Fig.~\ref{fig:chi_vs_mpi2}(a).  The
reanalysis is described in detail elsewhere \cite{ref:reducevariance}.
In this work we are concerned with the taste-breaking artifact.

We see that correcting for staggered fermion artifacts removes roughly
half the scaling violation in this range of pion masses.  The rest can
be attributed to the definition of the topological charge density
operator, including smoothing \cite{ref:topo.susc}.  It is plausible
that those corrections decrease as ${\cal O}(a^2)$.  We assume this
form in carrying out the rough continuum extrapolation.  The
extrapolation to the continuum starts by fitting the measured points
for each lattice spacing to the curve $1/\chi = c/m_{\pi,I}^2 + b$ and
then performing an extrapolation linearly in $a^2$ of the thus
smoothed values of $\chi$.  In the left panel the extrapolation was
done at constant Goldstone pion mass $m_{\pi}^2$; in the right panel,
at fixed $m_{\pi,I}^2$.  Because staggered fermion artifacts
contribute to scaling violations at fixed Goldstone pion mass, and
those artifacts decrease as ${\cal O}(\alpha_s^2 a^2)$, rather than
${\cal O}(a^2)$, the extrapolation in Fig~\ref{fig:chi_vs_mpi2}(a) is
somewhat less plausible than the extrapolation in
Fig~\ref{fig:chi_vs_mpi2}(b).

To make further progress we need simulations at a smaller lattice
spacing.

\section*{Acknowledgments}

C.D. and J.O. are grateful to their coauthors in
Ref.~\cite{ref:topo.susc} for helpful advice.  We thank particularly
Doug Toussaint.

\end{document}